\documentclass{pasj01}

\usepackage[small, bf]{caption}
\usepackage{threeparttable}
\usepackage{multirow}
\usepackage{color}

\Received{01-Dec-2017}
\Accepted{07-May-2018}
\Published{$\langle$publication date$\rangle$}

\begin{document}

\title{Discovery of recombining plasma from the faintest GeV SNR HB~21 and a possible scenario of the cosmic ray escaping from SNR shocks}

\author{Hiromasa Suzuki$^1$, Aya Bamba$^{1, 2}$, Kazuhiro Nakazawa$^{1, 2}$, Yoshihiro Furuta$^1$, Makoto Sawada$^{3, 4, 5}$, Ryo Yamazaki$^3$, and Katsuji Koyama$^6$}%

\altaffiltext{1}{Department of Physics, University of Tokyo, 7-3-1 Hongo, Bunkyo-ku, Tokyo 113-0033, Japan}
\altaffiltext{2}{Research Center for the Early Universe, School of Science, The University of Tokyo, 7-3-1 Hongo, Bunkyo-ku, Tokyo 113-0033, Japan}
\altaffiltext{3}{Department of Physics and Mathematics, Aoyama Gakuin University, 5-10-1 Fuchinobe, Chuo-ku, Sagamihara, Kanagawa 252-5258, Japan }
\altaffiltext{4}{X-ray Astrophysics Laboratory, NASA Goddard Space Flight Center, Greenbelt, MD 20771, USA}
\altaffiltext{5}{Department of Physics, University of Maryland Baltimore County, 1000 Hilltop Circle, Baltimore, MD 21250, USA}
\altaffiltext{6}{Department of Physics, Kyoto University, Kitashirakawa-oiwake-cho, Saikyo-ku Kyoto 606-8502, Japan}

\email{suzuki@juno.phys.s.u-tokyo.ac.jp}


\maketitle

\begin{abstract}
We present an X-ray study of the GeV gamma-ray supernova remnant (SNR) HB~21 with Suzaku. HB~21 is interacting with molecular clouds and the faintest in the GeV band among known GeV SNRs. We discovered strong radiative recombination continua of Si and S from the center of the remnant, which provide the direct evidence of a recombining plasma (RP). The total emission can be explained with the RP and ionizing plasma components. The electron temperature and recombination timescale of the RP component were estimated as 0.17~(0.15--0.18)~keV and 3.2~(2.0--4.8)\,$\times$\,10$^{11}~{\rm s~cm}^{-3}$, respectively.
The estimated age of the RP (RP age; $\sim 170~{\rm kyr}$) is the longest among known recombining GeV SNRs, because of very low density of electrons ($\sim 0.05~{\rm cm}^{-3}$).
{We have examined dependencies of GeV spectral indices on each of RP ages and SNR diameters for nine recombining GeV SNRs. Both showed possible positive correlations, indicating that both the parameters can be good indicators of properties of accelerated protons, for instance, degree of escape from the SNR shocks.}
A possible scenario for a process of proton escape is introduced; interaction with molecular clouds makes weaker magnetic turbulence and cosmic-ray protons escape, simultaneously cooling down the thermal electrons and generate an RP.

\end{abstract}

\section{Introduction}

Galactic cosmic rays are thought to originate from supernova remnant (SNR) shocks.
Very-high-energy gamma-ray studies revealed several SNRs accelerate protons to the energy beyond TeV (e.g., \cite{abdo}; \cite{ack}), by efficient diffusive shock acceleration with highly amplified magnetic field (e.g.,~\cite{bamba03}; \cite{vink}; \cite{yamazaki04}; \cite{bamba05}; \cite{bamba05b}).
Old GeV SNRs which interact with molecular clouds (MCs) show clear cutoff in their GeV spectra \citep{funk}, indicating that high-energy protons have already escaped from SNR shocks and become cosmic rays (e.g., \cite{ptuskin}; \cite{caprioli}; \cite{ohira10}; \cite{ohira}; \cite{ohira12}).
An important problem is, however, how accelerated protons escape from the amplified magnetic field in SNR shocks and become Galactic cosmic rays.

{Recent X-ray observations have found enhanced radiative recombination continua (RRCs) in the X-ray spectra of a dozen of middle-aged to old GeV SNRs. Such strong RRCs indicate that the plasmas are presently recombining rather than ionizing, and called recombining plasmas (RPs; e.g.,\,\cite{yamaguchi}; \cite{ozawa}; \cite{onishi}; \cite{sawada}). In such plasmas, the ionization temperature ($kT_{\rm z}$) is significantly higher than the electron temperature ($kT_\mathrm{e}$), implying that the plasma electrons have undergone a rapid cooling.
Since the plasmas have not reached collisional ionization equilibrium (CIE), the recombination timescale ($n_\mathrm{e}t$; elapsed time of relaxation due to Coulomb collision after the rapid cooling) must be small enough.}
The origin of RPs in SNRs is still unclear, but it is interesting that both the presence of soft GeV emission and RPs may occur in old and interacting SNRs.
Therefore, recombination of the plasma is a potential key to understand escaping of accelerated protons from SNR shocks.

Two principal scenarios have been proposed for the origin of RPs.
One is the rarefaction scenario \citep{itomasai}. If a supernova explosion occurs in a dense circumstellar matter (CSM), the CSM are quickly shock-heated and ionized. When the shock breaks out of the CSM and enters the low-density interstellar medium (ISM), $kT_{\rm e}$ is decreased rapidly by adiabatic cooling and an RP can be generated.
{The breakout timescale ($t_{\rm br}$) is estimated as
$t_{\rm br} \approx 10~(r_{\rm CSM}/0.1~{\rm pc})~(v_{\rm s}/10^4~{\rm km~s^{-1}})^{-1}~{\rm yr}$,
where $r_{\rm CSM}$ and $v_{\rm s}$ are the outer radius of the CSM and the shock velocity \citep{itomasai}.}
The other scenario to make an RP is the thermal conduction with MCs \citep{kawasaki}. When a shock collides with cold MCs, the plasma electrons can be rapidly cooled down and the plasma can become recombining.
{The thermal conduction timescale ($t_\mathrm{cond}$) is estimated to be
$t_\mathrm{cond} \approx 6 \times 10^{3}~(n_\mathrm{e}/1~\mathrm{cm}^{-3})~(l/1~\mathrm{pc})^2~(kT_\mathrm{e}/1~\mathrm{keV})^{-5/2}~\mathrm{yr}$,
where $n_\mathrm{e}$ and $l$ are the electron number density of the ISM and the temperature gradient scale length, respectively \citep{spitzer}. 
Note that \citet{cowie} suggests that the conduction timescale can be much longer due to the saturation of the heat flux.
In either case, the cooling from the initial temperature ($kT_{\rm init}$) should occur rapidly so that the plasma does not recover CIE at a new temperature. A characteristic timescale to reach CIE ($t_{\rm CIE}$) can be estimated as
$n_{\rm e}t_\mathrm{CIE} \approx \sum^{Z} _{z=0} (S_{z} + \alpha_{z})^{-1}$,
where $S_{z}$ and $\alpha_{z}$ are rate coefficients of ionization and recombination for an ion of the charge $z$ and the atomic number $Z$, respectively (\cite{masai}; \cite{smith}).
In the case of sulfur at a temperature of 1~keV,
$t_\mathrm{CIE} \approx 1.6 \times 10^4~(n_{\rm e}/1~\mathrm{cm}^{-3})^{-1}~\mathrm{yr}$ \citep{bryans}.
Thus {$t_{\rm br}$} or {$t_{\rm cond}$} should be shorter than {$t_{\rm CIE}$}, which can be satisfied, for example, if the CSM radius is sufficiently small or interacting MCs are clumpy enough to have small scale length, respectively.}

The SNR HB~21 is believed to be interacting with MCs (\cite{tatematsu}; \cite{koo}; \cite{byun}) and emits GeV gamma-rays \citep{pivato}.
The distance to HB~21 is estimated to be 1.7 kpc \citep{byun}.
It seems to have evolved much more than other GeV SNRs because it is very large ($\sim$ 100$'$ or $\sim$ 50 pc in diameter at 1.7~kpc), the faintest and softest in GeV band among known ones \citep{acero}. Therefore HB~21 must be a key to investigate possible physical relation between evolutionary stages and escaping processes of accelerated protons.
However, the properties of its thermal plasma has not yet been understood well.
\citet{lazendic} analyzed ASCA data and concluded that HB~21 had the CIE plasma, but statistics and energy resolution were limited.
We note the best-fit spectra show residuals alike RRCs of H-like Si and S, which implies that the plasma  may be recombining. Therefore we chose HB~21 as the best target to study escaping processes of accelerated protons by studying the thermal plasma, and observed it with Suzaku \citep{mitsuda}. With the low and stable background level and high accuracy calibration, Suzaku is the most suitable satellite for detailed investigation into the properties of the thermal plasma of HB~21, which is diffuse and faint in X-rays.

In this paper, we report Suzaku results of spectroscopy on HB~21, and comparison with other recombining GeV SNRs. Observation details are summarized in section 2. Results of the spectral analysis are presented in section 3, and discussed in section 4. 
In section 4, we also examine correlations between RPs and GeV properties among the known samples.
We summarize our results and discussions in section 5.
In this paper, we adopt 1.7 kpc as the distance to HB~21.

\section{Observations and Data Reduction}

\begin{table*}[t]
 \caption{Details of HB~21 and BGD observations.}
  \centering
  \begin{threeparttable}
   \begin{tabular}{l  l  c   l  l  c} \hline \hline
     Target name & Observation ID  & ({\it l, b})  & Start time & Stop time & Effective exposure (ks) \\ \hline
      
      HB~21  & 506005010 & ($88^\circ .8531$, $4^\circ.8048$) & 2011-04-09  & 2011-04-11 & 132.4    \\ 
      GRB 060105 & 900001010 & ($80^\circ .2506$, $9^\circ.9875$) & 2006-01-05 & 2006-01-06 & 42.2     \\ \hline
        \end{tabular}
        

        \end{threeparttable}

   \label{obs}
  
\end{table*}

Figure \ref{hb21} (a) shows WENSS 92 cm radio image and ROSAT PSPC 0.11--2.4 keV band X-ray image of HB~21\footnote{The data are available at $\langle$https://skyview.gsfc.nasa.gov/current/cgi/titlepage.pl$\rangle$. }.
We can see a clear shell in radio and center-filled emission in X-rays, which are typical characteristics of old SNRs (Mixed-Morphology; \cite{rho}).
We observed the central bright region of HB~21 with the Suzaku X-ray Imaging Spectrometer (XIS; \cite{koyama}).
X-ray emission from the source extends over the field of view (FoV) of the XIS, so the background region is not available from the same FoV.
Therefore we chose the data of the faintest and nearby point-like source GRB 060105 (hereafter, BGD) for the background estimation. Observation logs of HB~21 and BGD are shown in table~\ref{obs}.

In both observations, only three sets of onboard XIS0, 1, and 3 were operated. XIS1 is back-illuminated (BI) CCD whereas the others are front-illuminated (FI) ones.
The XIS were operated in the normal-clocking full-window mode.
The spaced-row charge injection (SCI; \cite{uchiyama}) technique was performed for all of the XIS in the HB~21 observation, but not for the BGD observation.
In the XIS data screening for both of the two observations, we eliminated the data acquired during the passage through the South Atlantic Anomaly (SAA), having elevation angles with respect to Earth's dark limb below 5$^{\circ}$, or with an elevation angle to the bright limb below 20${^\circ}$ in order to avoid contamination by emission from the bright limb.
%
We reprocessed the data with the calibration database version 2016-04-01.
The redistribution matrix files and the ancillary response files for the XIS were generated with {\tt xisrmfgen}, {\tt xissimarfgen} \citep{ishisaki}, respectively.
We generated ancillary response files with an assumption that the source region has an uniform brightness over a circular region of  \timeform{20'} in radius for both of the two observations.
For the spectral analysis, we used HEASoft 6.20, XSPEC 12.9.1, and AtomDB 3.0.9.

\section{Results}

\subsection{Images}

\begin{figure*}[htb]
\begin{center}
\includegraphics[width=16cm]{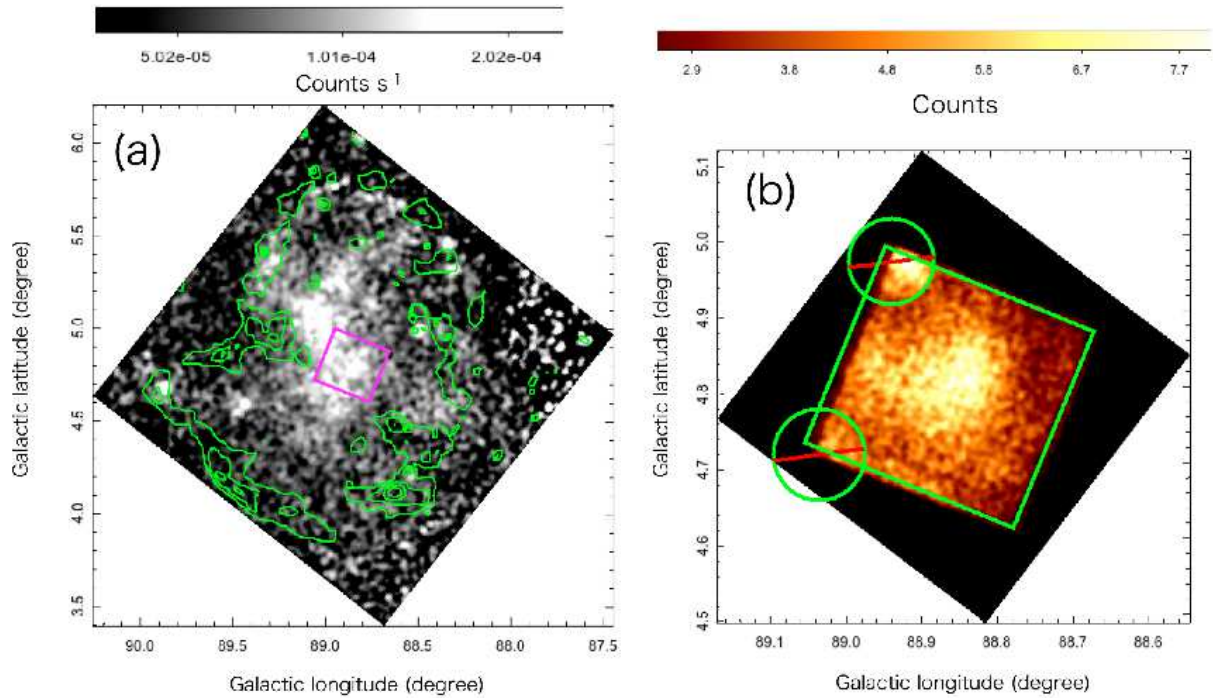}
\caption{ (a): ROSAT PSPC image (0.11--2.4 keV) of HB~21 smoothed with a Gaussian kernel of $\sigma = 1'$ (black and white). Overlaid green contours and magenta square represent WENSS 92 cm band image and the Suzaku XIS FoV, respectively. The image and contours are shown in the log scale. (b): The Suzaku XIS1 image (0.7--10 keV) of the center region smoothed with a Gaussian kernel of $\sigma = 0.4'$. The image is shown in the linear scale. Non-X-ray Background was not subtracted and no vignetting correction was performed. The source region for the spectral analysis is shown with a green square. The regions indicated by green circles are excluded from the analysis.}
\label{hb21}
\end{center}
\end{figure*}

\begin{figure}[htb]
\begin{center}
\includegraphics[width=8cm]{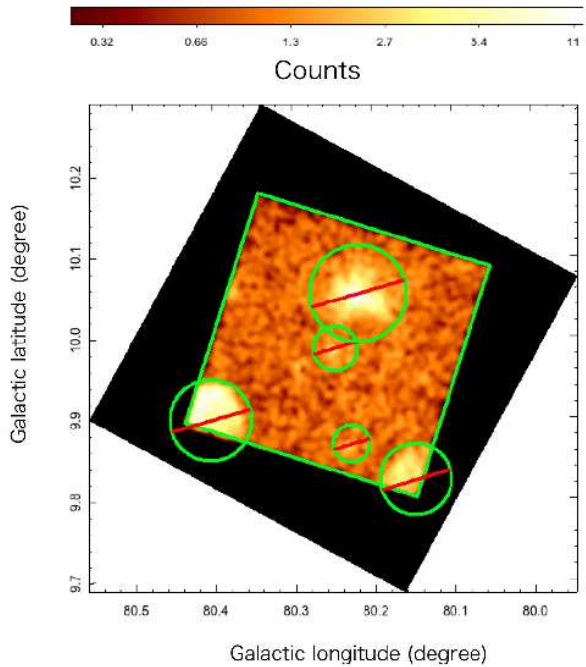}
\caption{The Suzaku XIS1 image (0.7--10 keV) of BGD observation smoothed with a Gaussian kernel of $\sigma = 0.4'$. The image is shown in the log scale. Non-X-ray Background was not subtracted and no vignetting correction was performed. The region for the spectral analysis is shown with a green square. The regions indicated by green circles are excluded from the analysis.}
\label{bgd}
\end{center}
\end{figure}

Figure \ref{hb21} (b) shows the Suzaku XIS1 0.7--10 keV band image of HB~21.
One can see that source emission extends all over the FoV.
The Suzaku XIS1 0.7--10 keV band image of the BGD observation is shown in figure~\ref{bgd}.
In this image, we can see several faint point-like sources.
The source-free region enclosed by green lines is used for the background estimation.


\subsection{Background spectra}

\begin{figure}[htb]
\begin{center}
\includegraphics[width=8cm]{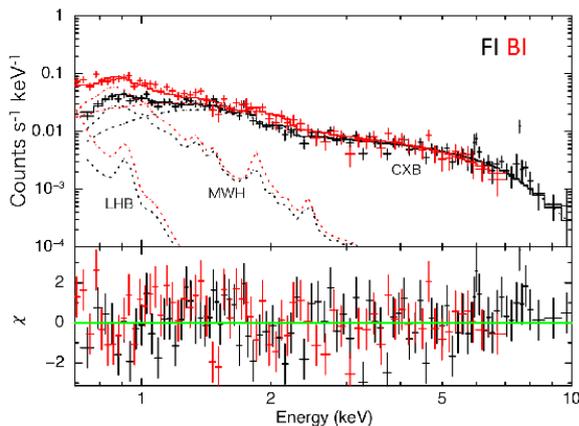}
\caption{ NXB-subtracted BGD spectra with the best-fit model and residuals. Dotted lines represent LHB, MWH and CXB whereas solid lines represent the sum of them. 
}
\label{bgdfit}
\end{center}
\end{figure}

We estimated the background spectrum in the HB~21 region with the BGD observation.
We assumed four components as the background emission, as with \citet{masui}; Non-X-ray Background (NXB), Local Hot Bubble (LHB; \cite{yoshino}), Milky Way Halo (MWH; \cite{masui}), and Cosmic X-ray Background (CXB; \cite{kushino}).
LHB and MWH are the emission from CIE plasmas surrounding the solar system ($\sim$\,0.1~keV) and our Galaxy ($\sim$\,0.8~keV), respectively.
We estimated the intensity of each component as follows.
The NXB spectrum was estimated by {\tt xisnxbgen} \citep{tawa}.
LHB and MWH were represented with CIE plasma models ({\tt apec}).
The electron temperature $kT_\mathrm{e}$ of the LHB model was fixed at 0.1 keV, whereas that of the MWH model was treated as a free parameter.
We assumed that CXB has the power-law spectrum with the photon index of 1.4 \citep{kushino}.
The normalizations of LHB, MWH, and CXB were treated as free parameters.
The Galactic absorption was commonly applied to the MWH and CXB components using the {\tt phabs} model \citep{balucinska} and the column density ($N_{\rm H}$) was fixed at $1.8 \times 10^{21}$~cm$^{-2}$~\citep{dickey}.
Then we fitted the NXB-subtracted BGD spectra with a model composed of LHB, absorbed MWH and CXB.
The model fitted the spectra well with $\chi^2 /\mathrm{d.o.f.} = 209 /161$.
The best-fit model and parameters are shown in figure~\ref{bgdfit} and table~\ref{bgdtab}, respectively.
The MWH flux obtained is the same order of magnitude as that in \citet{masui}.
The CXB flux is roughly consistent with the value in \citet{kushino} within $\approx$\,23\%.


\begin{table}[htb]
 \caption{Best-fit parameters of BGD spectra.}
  \centering
  \begin{threeparttable}
   \begin{tabular}{ p{15em}  l }
   \hline \hline Parameter & Value\tnote{*}\\
     \hline
     
     {\it kT}$_\mathrm{LHB}$ (keV) & 0.10 (fixed) \\ 
     {\it EM}$_{\mathrm{LHB} } (10^{-2}$) \tnote{\dag} &  1.2 (0.010--2.4) \\
     {\it kT}$_{\mathrm{MWH}}$ (keV) &   0.75 (0.71--0.79)  \\
     {\it EM}$_{\mathrm{MWH} } (10^{-4}$) \tnote{\dag} &  6.3 (5.5--7.0) \\
     
     norm$_{\mathrm{CXB} } (10^{-3}$) \tnote{\ddag}  & 1.28 (1.23--1.32) \\ 

     $\chi^2 /\mathrm{d.o.f.}$ & 209 /161 \\ \hline
     
   \end{tabular}
 
 \begin{tablenotes}\footnotesize
  \item[*]Errors indicate single parameter 90\% confidence level.
  \item[\dag] Emission measure in units of 10$^{-14}$(4$\pi${\it D}$^2$)$^{-1}$$\int${\it n}$_\mathrm{e}${\it n}$_\mathrm{H}${\it dV}~cm$^{-5}$, where {\it D}, {\it n}$_\mathrm{e}$ and {\it n}$_\mathrm{H}$ stand for distance (cm), electron and hydrogen number densities (cm$^{-3}$), respectively.  
  \item[\ddag]Normalization of {\tt powerlaw} model in units of cm$^{-2}$ s$^{-1}$ keV$^{-1}$ at 1 keV.
  
  \end{tablenotes}
  \end{threeparttable}
  \label{bgdtab}
  
\end{table}

\subsection{Source spectra}

\begin{figure}
\begin{center}
\includegraphics[width=8cm]{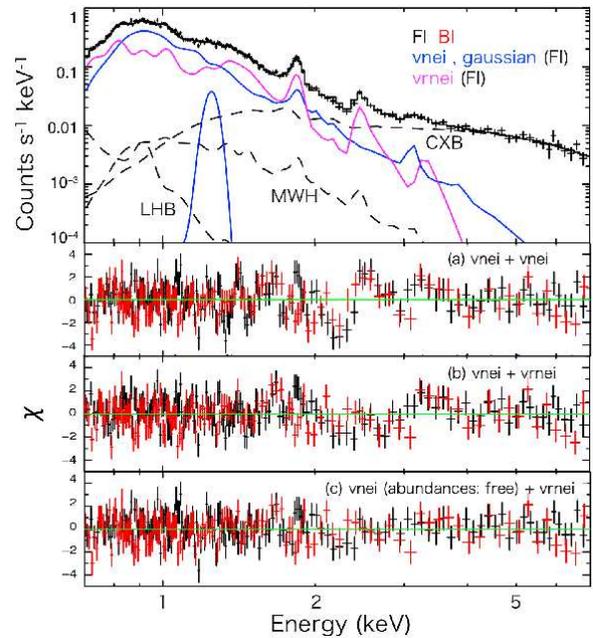}
\caption{NXB-subtracted HB~21 spectra of FI fitted with the model (c), and the residuals of FI and BI with the models (a), (b), and (c).
The solid lines represent {\tt vnei}, {\tt vrnei}, {\tt gaussian}, and the sum of all components. 
The dashed lines represent LHB, MWH, and CXB.
}
\label{hb21fit}
\end{center}
\end{figure}

We show the NXB-subtracted HB~21 spectra in figure~\ref{hb21fit}.
The spectra are much brighter than the background spectra estimated with BGD observation, thus we can assume that X-ray emission is mainly from HB~21.
Fe-L line complex, K-shell lines of He-like Si and S are clearly detected at $\approx $1, 1.9, and 2.5~keV, respectively.
Also, a hint of Ar K-shell line is found at 3.1~keV. The K-shell line of S has consistent energy centroid being He$\alpha$, and accompanied by smooth tail features at the higher energy side. Such feature is reminiscent of RRC \citep{yamaguchi}.

For the background emission, we used the model established with the BGD spectra but with different relative normalizations as described below.
We assumed the LHB flux in the HB~21 region is the same as that in the BGD region, and the MWH flux is different, since these two observations have different Galactic latitude by $\sim 5^\circ$.
In order to determine the MWH flux, we compared ROSAT PSPC (0.11--2.4 keV) fluxes around HB~21 to that in the BGD region, and obtained the flux ratio of HB~21:BGD = 0.80:1.0. Thus the MWH flux in HB~21 region was fixed at 0.8 times that obtained with the BGD observation.
The absorption column density was fixed at $7.6 \times 10^{21}$~cm$^{-2}$~\citep{dickey}.
Here, the MWH flux is only $\sim 0.8\%$ of the total emission from HB~21.
The CXB flux was treated as a free parameter.

As the emission from HB~21, we applied an optically thin thermal emission affected by Galactic absorption, which is added to the background component to fit the total spectra.
We applied one-$kT_\mathrm{e}$ ionizing plasma (IP; {\tt vnei}), CIE ({\tt vapec}), and RP ({\tt vrnei}; initial temperature $kT_{\mathrm{init} }$ was fixed at 2 keV) models, as the first step of the spectral fitting.
The abundances of Ne, Mg, Si, S, Ar, and Fe were treated as free parameters.
We added a line ({\tt gaussian} model) at $\approx$1.2 keV to supplement insufficient flux of Fe-L lines in the plasma models \citep{yamaguchi2}. The energy centroid and normalization were treated as free parameters.
All of the three models above did not represent the spectra with $\chi^2/\mathrm{d.o.f.}$\,=\,598/269, 613/270, and 659/269, respectively.
The one-$kT_\mathrm{e}$ models cannot explain the complicated structure below 2 keV.

We thus applied two-component IP (abundances: solar) + IP (abundances: free) model as shown in figure \ref{hb21fit}~(a), but it also did not explain the spectra with $\chi^2/\mathrm{d.o.f.}$\,=\,464/266.
The fit still shows large, negative residuals at\,$\sim$\,2.2 keV, requiring much lower bremsstrahlung continua.
In addition, it shows bump-like residuals at\,$\approx$\,2.7 and 3.5~keV, which correspond to K-shell ionization potential energies of Si and S, respectively. These features indicate that the spectra require an RP component which exhibits lower bremsstrahlung continuum with strong RRC emission of Si and S.
Therefore we applied IP (abundances: solar) + RP (abundances: free; $kT_{\mathrm{init} }$\,=\,free) model, and it greatly improved the fit (figure~\ref{hb21fit}~(b)), as the RP component shows high RRC emission of Si and S, and lowers the bremsstrahlung emission.
$\chi^2 /\mathrm{d.o.f.}$ was reduced to 358/265.
The best-fit parameters are shown in table \ref{hb21fit_tab}.
The CXB flux obtained is roughly consistent with the value in \citet{kushino} within $\approx$\,23\%.

This model still leaves residuals at\,$\approx$\,2.2, and 3.5~keV, which require much lower bremsstrahlung emission and higher RRC emission of S, respectively. These characteristics suggest much higher abundances of metals in the RP component, but they are prevented by fixed-abundance IP component. In particular, the abundance of S in the RP component cannot be higher because of the fixed strength of the K-shell line emission of S in the IP component.
Thus, we set metal abundances of the IP component to be free parameters.
This model further improved the fit with $\chi^2 /\mathrm{d.o.f.}$\,=\,325/259 (figure \ref{hb21fit} (c)), and it requires rather high abundances of Si, S, and Fe in the RP component, as expected.
The model (c) is statistically preferable to the model (b) with the F-test significance of 3.7~$\sigma$.
The best-fit parameters are shown in table \ref{hb21fit_tab2}.
The CXB flux again roughly agreed with the value in \citet{kushino} within $\approx$\,22\%.

   \begin{table*}[]
 \caption{ Best-fit parameters of the model (b) 
for HB~21 spectra.}
  \centering
  
  \begin{threeparttable}
  
   \begin{tabular}{p{12em}   l  l  l} \hline \hline
     Parameter\tnote{*} & {\tt vnei} & {\tt vrnei}  &  \\ \hline

     {\it N}$_\mathrm{H}$ ($10^{21}$ cm$^{-2}$) & --- & --- &  2.8 (2.6--3.0) \\
     \\
     {\it kT}$_\mathrm{e}$ (keV) & 0.788 (0.781--0.805) &  0.114 (0.110--0.119) & ---\\
     {\it kT}$_{\mathrm{init}}$ (keV) & --- & $>$ 7.7 & ---\\
         
     Ne (solar) & 1 (fixed) & 0.38 (0.31--0.49)  & ---\\
     Mg (solar) & 1 (fixed) & $<$ 0.066  & ---\\
     Si (solar) & 1 (fixed) & 0.89 (0.73--1.1) &  ---\\
     S (solar) & 1 (fixed) & 0.32 (0.13--0.51) & ---\\
     Ar (solar) &  1 (fixed) & 2.1 (0.77--3.5)  & --- \\
     Fe (solar) & 1 (fixed) & 11 (9.1--16)  & --- \\
     
     {\it EM} ($10^{-3}$) \tnote{\dag} & 6.2 (5.6--6.8) &  74 (53--96)  & --- \\
          
     {\it $n_\mathrm{e}t$} ($10^{11}$ s cm$^{-3}$) & 3.8 (3.3--4.4) & 4.8 (4.2--5.1)  & --- \\


\\    
     norm$_\mathrm{CXB}$ ($10^{-4}$) \tnote{\dag} & --- & --- & 8.3 (8.0--8.7) \\ 
  
\\  
  $\chi^2 /\mathrm{d.o.f.}$ & --- & --- &  358/265 \\ \hline
  
   \end{tabular}

   \begin{tablenotes}\footnotesize
   
   \item[*] Errors indicate single parameter 90\% confidence level.
   \item[\dag] Definitions are the same as Table \ref{bgdtab}.

   \end{tablenotes}
   \end{threeparttable}
   
  \label{hb21fit_tab}
\end{table*}

   \begin{table*}[]
 \caption{ Best-fit parameters of the model (c) 
for HB~21 spectra.}
  \centering
  
  \begin{threeparttable}
  
   \begin{tabular}{p{12em}   l   l  l} \hline \hline
     Parameter\tnote{*} & {\tt vnei} & {\tt vrnei}  &  \\ \hline

     {\it N}$_\mathrm{H}$ ($10^{21}$ cm$^{-2}$) & ---  & --- &  3.4 (3.0--3.8) \\
     \\
     {\it kT}$_\mathrm{e}$ (keV) & 0.80 (0.79--0.82) &  0.17 (0.15--0.18) & ---\\

     {\it kT}$_{\mathrm{init}}$ (keV) & --- & 0.58 (0.51--0.67) & ---\\
         
     Ne (solar) &   $<$ 0.89 & 3.6 (2.0--16) & ---\\
     Mg (solar) & $<$ 0.21 & 2.3 (1.7--4.3) & ---\\
     Si (solar) & 0.77 (0.49--1.1) & 23 (16--160) & ---\\
     S (solar) & $<$ 0.40 & $>$ 37 & ---\\
     Ar (solar) & 3.0 (1.3--4.8) & $<$ 140 & --- \\
     Fe (solar) & 1.4 (1.0--1.8) & 17 (7.3--110) & --- \\
     
     {\it EM} ($10^{-3}$) \tnote{\dag} & 6.7 (5.1--8.8) &  22 (7.5--38) & --- \\
          
     {\it $n_\mathrm{e}t$} ($10^{11}$ s cm$^{-3}$) & $>$ 4.2 & 3.2 (2.0--4.8) & --- \\


\\    
     norm$_\mathrm{CXB}$ ($10^{-4}$) \tnote{\dag} & --- & --- & 8.4 (8.0--8.6) \\ 
  
\\  
  $\chi^2 /\mathrm{d.o.f.}$ & --- & ---  &  324/259 \\ \hline
  
   \end{tabular}

   \begin{tablenotes}\footnotesize
   
   \item[*] Errors indicate single parameter 90\% confidence level.
   \item[\dag] Definitions are the same as Table \ref{bgdtab}.

   \end{tablenotes}
   \end{threeparttable}
   
  \label{hb21fit_tab2}
\end{table*}

\section{Discussion}

We discovered an RP from the old GeV SNR HB~21, for the first time.
Two plasma models (b and c) gave relatively good fits and both require an RP component.
In this section, we adopt the model (c) as the best-fit model.

\subsection{Physical properties of HB~21}

First we check the validity of the distance to HB~21.
We obtained the absorption column density ($N_\mathrm{H}$) of ($3.4\,\pm\,0.4) \times 10^{21}~\mathrm{cm}^{-2}$, which is consistent with the ASCA results (\cite{lee}; \cite{yoshita}; \cite{lazendic}). This supports the distance of 1.7 kpc derived by \citet{byun}, which was estimated by the relation between $N_\mathrm{H}$ and velocity-resolved intensities of H\,\emissiontype{I} and $^{12}$CO emission.
 
Next, we estimate the physical parameters of the plasma.
We found that HB~21 plasma consists of the IP and RP components.
The IP and RP components require roughly consistent with, and higher abundances than the solar value, respectively. These results suggest that the plasma has the ionizing ISM and recombining ejecta components.
The electron densities of the IP and RP are estimated using the emission measures as $\sim 0.033~f^{-0.5}~\mathrm{cm}^{-3}$ and $\sim 0.060~f^{-0.5}~\mathrm{cm}^{-3}$, respectively, assuming the HB~21 plasma to be an ellipsoid with the radii of 10~pc, 10~pc, and 15~pc (the volume of $\sim 2 \times 10^{59}~\mathrm{cm}^{3}$).
Here, $f$ is a volume filling factor of the plasma.
{
The shocked ISM density around HB~21 ($\sim 0.033~f^{-0.5}$~cm$^{-3}$) is much smaller than those of other SNRs.}
The plasma mass of the whole remnant {(including both the ISM and ejecta)} are estimated as $\sim 7.8~f^{0.5}~M_\odot$. We also estimate the age of the RP (hereafter, RP age) as $\sim 170$ kyr, with $n_{\rm e}$ and $n_{\rm e}t$.
Here, these two values are derived assuming that these values are constant throughout the evolution of the SNR.
The estimated RP age is the longest among known recombining SNRs due to very small value of $n_{\rm e}$, indicating that HB~21 is one of the most evolved recombining SNRs.

\subsection{Comparison among recombining GeV SNRs}

Here, we compare the properties of GeV SNRs which have RPs in order to search for a clue of particle escape.
We select 12 SNRs which are known to have both GeV emission and RPs as shown in table \ref{nt_gamma_tab}.
RP ages can be calculated for nine SNRs.
We basically adopt the values of GeV spectral indices presented in \citet{acero} to get an uniform sample, but as for some SNRs, we use other results since they are not included in \citet{acero}.
\textcolor{black}{
We note that Kes~17 and 3C 391 have results from other literatures as well (\cite{wu}; \cite{ergin391}). Here we use their values from \citet{acero}, unless mentioned otherwise.
}

Higher-energy protons easily escape from SNR shocks, since their mean free paths are longer than those of lower-energy ones {(\cite{ptuskin}; \cite{caprioli}; \cite{ohira10}; \cite{ohira}; \cite{ohira12}).}
Along with the progress of proton escape, GeV gamma-ray flux gradually decreases from higher energy band, resulting in softer GeV spectrum.
Thus, the spectral index in the GeV band should be a good indicator of the progress of proton escape.

Here, we make a hypothesis on how protons escape from SNR shocks; interaction with MCs causes weaker magnetic turbulence due to the wave damping occurred by ion-neutral collisions {(\cite{drury}; \cite{bykov})}, resulting in the escape of cosmic-ray protons, simultaneously cooling down the thermal electrons in the plasma and generate an RP (\cite{masai}; \cite{kawasaki}).
According to the scenario, the spectrum of GeV emission from an SNR gradually becomes soft after the collision with MCs, whereas the recombination-dominated state in the plasma gradually terminates since $kT_{\rm z}$ becomes closer to $kT_\mathrm{e}$ as the RP age increases.
\textcolor{black}{
Note that shock-cloud interactions and the subsequent proton escape happen at shock fronts, and plasmas are gradually cooled down from the outer to the central regions either by rarefaction or thermal conduction (\cite{itomasai}; \cite{kawasaki}; \cite{matsumura18}). Therefore, we expect that RPs which locate far inside the shock fronts can also be used to roughly estimate the progress of the proton escape.}
In the case of HB~21, the spectral index in the GeV band is the largest among GeV SNRs ($\sim$\,3.0; \cite{acero}), whereas we found that the RP age of the plasma is the longest among recombining SNRs (see table \ref{nt_gamma_tab}). These facts support the hypothesis.

\begin{figure}[htb]
\begin{center}
\includegraphics[width=8cm,]{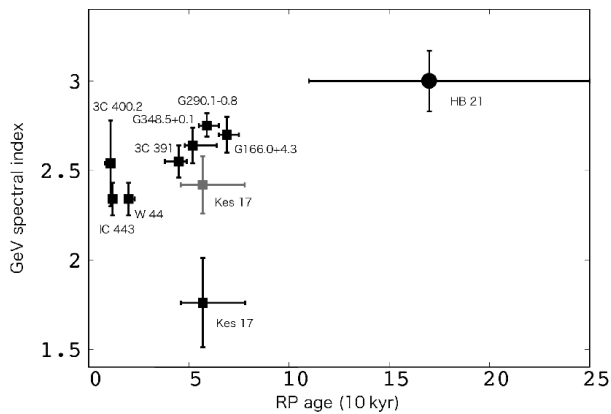}
\caption{Relation between RP ages and GeV spectral indices of recombining GeV SNRs. Two results of GeV spectral index of Kes~17 are shown (black: \citet{acero}, grey: \citet{wu}).}
\label{nt_gamma}
\end{center}
\end{figure}

{We check the correlation between GeV spectral indices and RP ages for nine SNRs (table \ref{nt_gamma_tab}).}
The result is shown in figure~\ref{nt_gamma} and table \ref{size_nt_tab}.
{A positive correlation with the correlation coefficient of 0.50\,$\pm$\,0.12 (1 $\sigma$ statistical error)\,$\pm$\,0.32 (standard error of the regression) is found.
This indicates that RP ages can possibly be used to estimate the progress of proton escape.
Note that we ignore the compression and re-acceleration of Galactic cosmic rays, which could affect the GeV emission from old SNRs (\cite{uchiyama10}; \cite{cardillo}).}

\begin{figure}[htb]
\begin{center}
\includegraphics[width=8cm,]{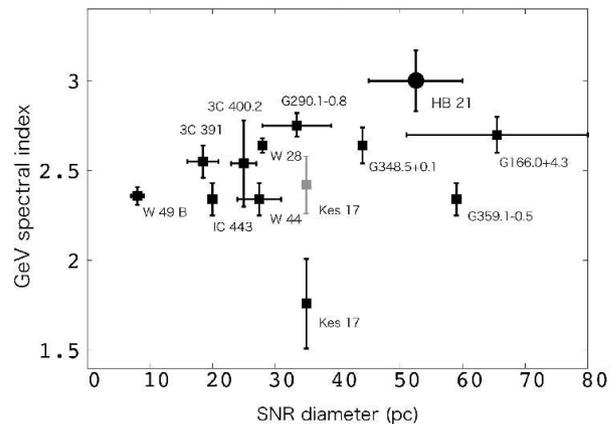}
\caption{Relation between SNR diameters and GeV spectral indices of recombining GeV SNRs. Two results of GeV spectral index of Kes~17 are shown (black: \citet{acero}, grey: \citet{wu}).}
\label{size_gamma}
\end{center}

\end{figure}

{The progress of proton escape can also be determined by the shock velocity, which is difficult to measure especially in cases of old SNRs.
On the other hand, SNR diameters, which reflect shock velocities, can be easily measured from radio images. If the progress of proton escape is determined by the shock velocity, we can expect a strong correlation between GeV luminosities and SNR diameters, or, GeV spectral indices and SNR diameters.}
\citet{bamba16} checked the former correlation, and found that GeV luminosities do not show significant decrease along with the diameters.
Here, we check the latter for 12 SNRs.
{As shown in figure \ref{size_gamma} and table \ref{size_nt_tab}, we find only a weak correlation with the coefficient of 0.28\,$\pm$\,0.07\,$\pm$\,0.30.
The result indicates that physical processes which are responsible for particle escape are not strongly related to SNR diameters, and as a result, shock velocities.
We note that, more detailed analysis is needed in the future, to estimate the shock velocities precisely, since they are affected by the kinetic energies of the supernova explosions and the ISM densities.}

{We note that the correlation results in both cases have large uncertainties than the statistical ones; the correlation between RP ages and GeV spectral indices could be supported only by HB~21. If we eliminate HB~21 from the sample, the correlation coefficient is reduced to 0.14\,$\pm$\,0.17\,$\pm$\,0.41.
Similarly, SNR diameters and GeV spectral indices have no correlation if we eliminate HB~21, with the correlation coefficient of 0.15\,$\pm$\,0.08\,$\pm$\,0.33.
These results are also shown in table \ref{size_nt_tab}.
In addition, our analysis uses parameters from various literatures which use different methods of analysis from each other.
Kes~17 and 3C 391 have other results of the GeV spectral indices (\cite{wu}; \cite{ergin391}), probably because of the differences in the energy range and sky region used for the spectral analyses, and the background estimation.
In particular, Kes 17 has very large discrepancy between the two results (see table \ref{nt_gamma_tab}). Therefore we check the correlations using the GeV spectral index of Kes~17 from \citet{wu} as shown in table \ref{size_nt_tab}. The results exhibit positive correlations with GeV spectral indices with the correlation coefficient of 0.80\,$\pm$\,0.10\,$\pm$\,0.19 (RP ages) and 0.41\,$\pm$\,0.09\,$\pm$\,0.28 (SNR diameters).}
{Taking account of these situations, we conclude that no clear difference is found in the significances of the two correlation results on RP ages and SNR diameters.
Nevertheless we would like to emphasize that the RP age can be a good parameter to estimate a timescale of the proton escape.}
In order to study the correlations precisely, as the future work, it will be needed that GeV spectral analysis for all of the recombining GeV SNRs with the uniform energy range, and X-ray analysis to estimate RP ages for all of the recombining SNRs.
Spatially-resolved analyses are also needed to investigate the generating processes of RPs in GeV SNRs.

\begin{table*}[h]
 \caption{Properties of 12 recombining GeV SNRs}
  \centering
  
  \begin{threeparttable}
  
   \begin{tabular}{l  l  l  l  l  l } \hline \hline
     Sample & SNR diameter (pc)\tnote{*}& $n_\mathrm{e}$~of RP (cm$^{-3}$)\tnote{\dag} & RP age (10 kyr)\tnote{\ddag} & GeV spectral index\tnote{\S} & References\tnote{$\|$}  \\ \hline
     W 49 B & 8 (7--9) & 4.5 & --- & 2.36 (2.31--2.41) & (1), (2), (3), (4), (5), (6) \\
     G359.1--0.5 & 59 & --- &  --- & 2.34 (2.25--2.43) & (1), (2), (7), (8) \\
     W 28 & 28  & 0.77 & --- & 2.64 (2.60--2.68) & (1), (2), (9), (10), (11), (12), (13) \\
     IC 443 & 20 & 1.7 & 1.2 (1.1--1.3) & 2.34 (2.25--2.43) & (1), (2), (14), (15), (16) \\
     3C 391 & 18.5 (16--21) & 0.9 & 4.5 (3.8--4.9) & 2.55 (2.46--2.64) & (1), (2), (17), (18)  \\
     3C 400.2 & 25 (23--27) & 0.51 & 1.1 (0.94--1.2) & 2.54 (2.30--2.78) & (1), (19), (20) \\
     G166.0+4.3 & 65.5 (51--80) & 0.28 & 6.9 (6.5--7.5) & 2.70 (2.60--2.80) & (1), (21), (22), (23) \\
     G290.1--0.8 & 33.5 (28--39) & 0.66 & 5.9 (5.5--6.5) & 2.75 (2.69--2.82) & (1), (24), (25), (26) \\
     G348.5+0.1 & 44 & 0.82 & 5.2 (4.8--6.4) & 2.64 (2.54--2.74) & (1), (2), (27) \\
     Kes 17 & 35 & 0.9 & 5.7 (4.6--7.8) & 1.76 (1.51--2.01) & (1), (2), (28), (29) \\
      & & & & 2.42 (2.26--2.58) & (30) \\
     W 44 & 27.5 (24--31) & 1 & 2.0 (1.8--2.3) & 2.34 (2.25--2.43) & (1), (2), (31), (32) \\
     HB~21 & 52.5 (45--60) & 0.060 &  17 (11--25) & 3.00 (2.83--3.17) & (1), (2), (33), this work \\ \hline
     

   \end{tabular}

   \begin{tablenotes}\footnotesize
   
   \item[*] Errors indicate the radii between major and minor axes.
   \item[\dag] The values presented for each object, but in the case of G290.1--0.8, we estimated it assuming the SNR plasma to be an ellipsoid since we have no information on the density of the RP in \citet{kamitsukasa}.
   \item[\ddag] Errors indicate single parameter 90\% confidence level. Only errors of $n_{\rm e}t$ are shown. The values of the central plasmas for G290.1-0.8, W 44, and HB~21, and the average values over the whole RPs for the others.
   \item[\S] Spectral index of the best-fit {\tt powerlaw} model for each GeV spectrum. Only statistical errors are shown.

   \item[$\|$] References: 
   (1) \citet{green}; (2) \citet{acero}; (3) \citet{ozawa}; (4) \citet{miceli}; (5) \citet{radhak_a}; (6) \citet{moffett};
   (7) \citet{onishi}; (8) \citet{uchida92};
   (9) \citet{sawada}; (10) \citet{zhou}; (11) \citet{ilovaisky}; (12) \citet{goudis}; (13) \citet{velazquez};
   (14) \citet{yamaguchi}; (15) \citet{matsumura2}; (16) \citet{welsh};
   (17) \citet{sato}; (18) \citet{radhak_a};
   (19) \citet{ergin}; (20) \citet{giacani};
   (21) \citet{matsumura}; (22) \citet{araya}; (23) \citet{landecker};
   (24) \citet{kamitsukasa}; (25) \citet{auchettl}; (26) \citet{reynoso};
   (27) \citet{yamauchi};
   (28) \citet{washino}; (29) \citet{caswell};
   (30) \citet{wu};
   (31) \citet{uchida12}; (32) \citet{claussen};
   (33) \citet{byun}
      
   \end{tablenotes}
   \end{threeparttable}
   
  \label{nt_gamma_tab}
\end{table*}

\begin{table*}[htb]
\caption{Correlation coefficients of the several cases}
\label{size_nt_tab}
\begin{center}
\begin{threeparttable}
  
\begin{tabular}{p{9em}  p{4em}  p{5em}  l  l } \hline \hline
Case & Include HB~21 & Reference of Kes 17\tnote{*} & Correlation coefficient\tnote{\dag} & Significance ($\sigma$)\tnote{\ddag} \\
\hline
\multirow{2}{8.5em}{RP ages and\,\,\,\,\,\, GeV spectral indices} & Yes & A & $0.50 \pm 0.12 \pm 0.32$ & 1.5 \\
&  & W & $0.80 \pm 0.10 \pm 0.19$ & 3.7 \\
&  & N & $0.85 \pm 0.09 \pm 0.17$ & 4.4 \\
& No & A & $0.14 \pm 0.17 \pm 0.41$ & 0.32 \\
&  & W & $0.62 \pm 0.19 \pm 0.30$ & 1.7 \\
&  & N & $0.77 \pm 0.24 \pm 0.16$ & 2.7  \\
\hline
\multirow{2}{8.5em}{Diameters and GeV spectral indices} & Yes & A & $0.28 \pm0.07 \pm 0.30$ & 0.91 \\
&  & W & $0.41 \pm 0.09 \pm 0.28$ & 1.4 \\
&  & N & $0.42 \pm 0.09 \pm 0.30$ & 1.3 \\
& No & A & $0.15 \pm 0.08 \pm 0.33$ & 0.44 \\
&  & W & $0.29 \pm 0.11 \pm 0.32$ & 0.86 \\
&  & N & $0.30 \pm 0.33 \pm 0.11$  &0.86  \\
\hline

\end{tabular}

\begin{tablenotes}
\footnotesize

\item[*] A: GeV spectral index from \citet{acero}; W: GeV spectral index from \citet{wu}; N: without Kes~17.
\item[\dag] Value $\pm$ (1 $\sigma$ statistical error) $\pm$ (standard error of the regression).
\item[\ddag] Significance of positivity of the correlation coefficient.

\end{tablenotes}
\end{threeparttable}
\end{center}
\end{table*}

\section{Conclusion}

We conducted an X-ray analysis on the faintest GeV SNR HB~21.
A strong Si- and S-RRC (H-like) were discovered in the spectra, which provide the direct evidence of the RP component ($kT_\mathrm{e}$\,=\,0.17~(0.15--0.18)~keV, $n_\mathrm{e}t$\,=\,3.2~(2.0--4.8)\,$\times\,10^{11}$~s~cm$^{-3}$) besides the IP component ($kT_\mathrm{e}$\,=\,0.80~(0.79--0.82)~keV, $n_\mathrm{e}t\,>\,4.2\,\times\,10^{11}$~s~cm$^{-3}$).
The estimated RP age ($\sim$\,170~kyr) is the longest among known recombining SNRs.
We also compared GeV spectral indices to RP ages and SNR diameters for nine recombining GeV SNRs.
{
Both parameter sets showed possible positive correlations, and we found no clear difference between them in the significances of the correlations.
This indicate that both the RP age and SNR diameter can be good parameters to estimate the progress of proton escape from SNR shocks, and in particular, the RP age possibly indicates a timescale of the escape.}
A possible scenario for a process of the escape is introduced; interaction with MCs makes weaker magnetic turbulence and cosmic-ray protons escape, simultaneously cooling down the thermal electrons and generate an RP.

\begin{ack}

The authors deeply appreciate Suzaku team members.
We are especially grateful to Koji Mori for helpful advice on the XIS calibration.
We also thank the anonymous referee for valuable comments, especially on the correlation analysis.
This work was supported by Grants-in-Aid for Scientific Research from the Ministry of Education, Culture, Sports,
Science and Technology (MEXT) of Japan, No. 15K05107 and 15K05088.
This work was also supported by a grant from the Hayakawa Satio Fund awarded by the Astronomical Society of Japan.

\end{ack}

\end{document}